\documentclass[twocolumn]{aastex631}
\usepackage{CJKutf8} 
\usepackage{amsmath}
\usepackage{comment}
\usepackage{physics}
\usepackage{verbatim}

\usepackage{longtable}

\newcommand\arsec{$^{\prime\prime}$}

\usepackage{soul}

\submitjournal{ApJL}

\begin{document}

\title{LIGHTS. The Thin Encircling Stellar Stream of NGC 3938}
  
\correspondingauthor{Dennis Zaritsky}
\email{dennis.zaritsky@gmail.com}

\author[0000-0002-5177-727X]{Dennis Zaritsky}
\affiliation{Steward Observatory and Department of Astronomy, University of Arizona, 933 N. Cherry Ave., Tucson, AZ 85721, USA}

\author[0000-0001-8042-5794]{Jacob Nibauer}
\affiliation{Department of Astrophysical Sciences, Princeton University, 4 Ivy Ln, Princeton, NJ 08544, USA}

\author[0009-0001-2377-272X]{Giulia Golini}
\affiliation{School of Physics, University of New South Wales, NSW 2052, Australia}

\author[0009-0003-6502-7714]{Ignacio Ruiz Cejudo}
\affiliation{Instituto de Astrof\'isica de Canarias, c/ V\'ia L\'actea s/n, 38205 La Laguna, Tenerife, Spain}
\affiliation{Departamento de Astrof\'isica, Universidad de La Laguna, 38206 La Laguna, Tenerife, Spain}


\author[0000-0001-8647-2874]{Ignacio Trujillo}
\affiliation{Instituto de Astrof\'isica de Canarias, c/ V\'ia L\'actea s/n, 38205 La Laguna, Tenerife, Spain}
\affiliation{Departamento de Astrof\'isica, Universidad de La Laguna, 38206 La Laguna, Tenerife, Spain}

\author[0000-0003-0256-5446]{Sarah Pearson}
\affiliation{DARK, Niels Bohr Institute, University of Copenhagen, Jagtvej 155A, 2200 Copenhagen,  Denmark}


\author[0000-0002-1598-5995]{Nushkia Chamba}
\affiliation{
NASA Ames Research Center, Space Science and Astrobiology Division M.S. 245-6, Moffett Field, CA 94035, USA}

\author[0000-0003-2069-9413]{Chen-Yu Chuang}
\affiliation{Steward Observatory and Department of Astronomy, University of Arizona, 933 N. Cherry Ave., Tucson, AZ 85721, USA}

\author[0000-0001-8647-2874]{Mauro D'Onofrio}
\affiliation{Department of Physics and Astronomy, University of Padua, Vicolo Osservatorio 3, 35122 Padova, Italy}

\author[0000-0002-6672-1199]{Sepideh Eskandarlou}
\affiliation{Centro de Estudios de F\'isica del Cosmos de Arag\'on (CEFCA), Plaza San Juan, 1, E-44001, Teruel, Spain}

\author[0009-0001-7407-2491]{Sergio Guerra Arencibia}
\affiliation{Instituto de Astrof\'isica de Canarias, c/ V\'ia L\'actea s/n, 38205 La Laguna, Tenerife, Spain}
\affiliation{Departamento de Astrof\'isica, Universidad de La Laguna, 38206 La Laguna, Tenerife, Spain}

\author[0000-0003-3449-2288]{S.Zahra Hosseini-ShahiSavandi}
\affiliation{Department of Physics and Astronomy, University of Padua, Vicolo Osservatorio 3, 35122 Padova, Italy}


\author[0009-0007-7712-0683]{Ouldouz Kaboud}
\affiliation{Department of Physics and Astronomy, University of Padua, Vicolo Osservatorio 3, 35122 Padova, Italy}

\author[0009-0003-0674-9813]{Minh Ngoc Le}
\affiliation{Instituto de Astrof\'sica de Canarias, c/ V\'ia L\'actea s/n, 38205 La Laguna, Tenerife, Spain}
\affiliation{Departamento de Astrofísica, Universidad de La Laguna, 38206 La Laguna, Tenerife, Spain}

\author[0000-0003-2939-8668]{Garreth Martin}
\affiliation{School of Physics and Astronomy, University of Nottingham, University Park, Nottingham NG7 2RD, UK}

\author[0000-0001-7847-0393]{Mireia Montes}
\affiliation{Institute of Space Sciences (ICE, CSIC), Campus UAB, Carrer de Can Magrans, s/n, 08193 Barcelona, Spain}

\author[0000-0001-9000-5507]{Samane Raji}
\affiliation{Departamento de F\'isica Te\'orica, At\'omica, y \'Optica, Universidad de Valladolid, 470111, Valladolid, Spain}
\affiliation{ Laboratory for Disruptive Interdisciplinary Science (LaDIS), Universidad de Valladolid, 47011 Valladolid, Spain}

\author[0000-0002-3849-3467]{Javier Román}
\affiliation{Departamento de Física de la Tierra y Astrofísica, Universidad Complutense de Madrid, E-28040 Madrid, Spain}
\affiliation{Departamento de Física, Universidad de Córdoba, Campus de Rabanales, Edificio Albert Einstein, E-14071 Córdoba, Spain}

\author[0009-0001-9574-8585]{Nafise Sedighi}
\affiliation{Instituto de Astrof\'sica de Canarias, c/ V\'ia L\'actea s/n, 38205 La Laguna, Tenerife, Spain}
\affiliation{Departamento de Astrofísica, Universidad de La Laguna, 38206 La Laguna, Tenerife, Spain}

\author[0009-0004-5054-5946]{Zahra Sharbaf}
\affiliation{Instituto de Astrof\'sica de Canarias, c/ V\'ia L\'actea s/n, 38205 La Laguna,
Tenerife, Spain}
\affiliation{Departamento de Astrofísica, Universidad de La Laguna, 38206 La Laguna, 
Tenerife, Spain}

\begin{abstract}
We present a stellar stream found in images of the nearby, nearly face-on, late-type galaxy, NGC 3938 obtained for the 
LBT Imaging of Galactic Halos and Tidal Structures
(LIGHTS) survey that is thin, has very low mean surface brightness
($\langle\mu_g\rangle \approx$ 28.7 mag arcsec$^{-2}$ and $\langle\mu_r\rangle \approx$ 28.1 mag arcsec$^{-2}$), appears
to lie nearly on the plane of the sky, and wraps more
than half way around a host galaxy that is otherwise
apparently isolated. We estimate that the progenitor had a stellar mass of $\sim 3.7\times 10^7$ M$_\odot$. Despite an intriguing apparent offset between the centroid of the host galaxy and the apparent center of the stream orbit, we find that we can reproduce the morphology, including this apparent off-centering, with simple models and standard assumptions about the host (thin disk centered within a canonical spherical dark matter halo) and the progenitor satellite orbit. We identify a number of detailed features of the stream, such as changes in curvature and density, that will require more complex models to reproduce. Even this rather simple system provides a rich set of constraints with which to explore the accretion history and gravitational potential of an otherwise unremarkable late-type galaxy.
Given the depth of the LIGHTS images, this system is an example of the types of stellar stream that could be found in a majority of nearby giant galaxies with the 10-year stack of Rubin/LSST data.
\end{abstract}


\section{Introduction}
\label{sec:intro}

As imaging pushes to lower and lower surface brightness limits, nearly every galaxy image is expected to reveal distinguishable relics of the galaxy's hierarchical growth above any smooth stellar halo that may be present \citep[e.g.,][]{cooper,martin}. 
The LBT Imaging of Galactic Halos and Tidal Structures
(LIGHTS) survey \citep{LIGHTSs,lights-z}, which is deeper than the expected depth of the 10-year Rubin/LSST stack\footnote{The estimated 3$\sigma$ surface brightness limit for a 10\arsec$\times$10\arsec\ box 
in the Rubin/LSST stack is 30.3 mag arcsec$^{-2}$ for both $g$ and $r$ bands 
(https://community.lsst.org/t/limiting-surface-brightness/6214).}
, provides a testing ground for this prediction and an opportunity to identify tidal debris from low mass satellite galaxies.
We present a discovery of a stellar stream that nearly encircles a nearby, face-on, late-type galaxy. For various reasons that we will describe, this system is potentially a valuable probe of this galaxy's accretion history and dark matter halo.

At the current time, we are well into what can be defined as the third stage of the study of galactic dark matter halos. The first stage began with empirical evidence for dark matter in galaxies \citep{babcock,rubin,roberts} and ended with the conclusive demonstration that dark matter halos extend up to and beyond the virial radii of galaxies and with simple estimates (``factor of two") of their masses \citep{zaritsky,breinerd,fischer,prada}. The second stage, following but also somewhat concurrent, developed the theoretical framework for structure and halo growth \citep[cf.][]{frenk}, a deeper understanding of the universal structure of such halos \citep{dubinski,nfw}, and demonstrated the centrality of halo growth to galaxy formation and evolution \citep{white,blumenthal,behroozi19}. The third stage, which we are currently well into, is where detailed halo properties are used as tools with which to further our understanding of the fundamental nature of dark matter. Such efforts include, but are not limited to, studies of halo inner density profiles \citep{moore,kravtsov,read}, gravitational wakes induced by dynamical friction \citep{furlanetto,garavito19,foote2023}, and gaps and other structures in cold stellar streams \citep[for a recent review, see][]{bonaca}.

The study of galactic halo stellar streams is particularly promising in this regard because, for certain questions, one sidesteps having to model the central region of galaxies where baryonic effects can dominate \citep{blumenthal86,bullock17}. Streams are 
a sensitive probe of both the global potential \citep{johnston99,buist15,bonaca-hogg} and any local perturbation produced by dark matter substructure  \citep{ibata, johnston2002}.
Quantitative results have come primarily from streams within the Galactic halo, for which there is now 6-D phase space information \citep{bonaca}, but our galaxy is a single example and its halo is strongly influenced by the current infall of the Magellanic Clouds \citep{garavito19,conroy21}. Streams in galaxies outside of the Local Group provide additional opportunities \citep[cf.,][]{shang,martinez-delgado,roman,sola2025}, although for those we generally only have 2-D information, which makes it more difficult to construct definitive models.

Recent work shows how to recover information about the host galaxy’s halo and the stream progenitor using only deep imaging \citep{amorisco, nibauer, walder, nibpear2025}. However, given the incomplete information we have for extragalactic streams, it is unclear whether it is more fruitful to conduct a broad statistical study of such streams \citep{kado-fong,morales,martinez-delgado23,martin, miro2024, miro2025} or to find specific, well-chosen examples that are, for one reason or another,  highly constrained \citep[e.g.][]{mcconnachie2009,denja2016,denja2019,mueller2025} and simpler to model \citep{fardal2013,Pearson}. Experience suggests that we exploit both categories of study.

We present the discovery of a stream in the latter category that may prove particularly useful. First, it is a stream around a nearly face-on galaxy (NGC 3938), which helps to define one of the likely principal planes of the system. Second, the galaxy has no luminous companions\footnote{A search of the NASA Extragalactic Database yields no galaxies that lie within a projected separation of 300 kpc, a recessional velocity window of 500 km s$^{-1}$, and an $r$ band magnitude difference that is $<$ 5 mag.} and a late-type morphology, suggesting that the halo is as undisturbed as possible within a hierarchical picture of galaxy formation. Third, the stream is of low surface brightness, suggesting that the progenitor itself was not sufficiently massive to significantly disturb NGC 3938's halo. Fourth, the stream extends over 180$^\circ$ around the galaxy, allowing us to trace the majority of the most recent orbit. Fifth, the stream traces a roughly circular path around the galaxy, suggesting that it too lies in the nearly face-on disk plane, unless its orbital plane just happens to be oriented at the proper angle to foreshorten an elliptical orbit into an apparently circular one. 

Our Large Binocular Telescope (LBT) Imaging of Galactic Halos and Tidal Structures (LIGHTS) survey is motivated by a desire to establish the range of stellar halo properties and test the basics of the galaxy formation scenario \citep{LIGHTSs,lights-z}.
We aim to provide the best possible measurements of the diffuse halo stellar light for as large a sample of nearby galaxies as possible. We therefore aim to reach as faint a surface brightness limit as possible while retaining the highest possible angular resolution with which to classify and mask contaminating sources. In the case of NGC 3938, an unbarred late-type galaxy \citep[SA(s)c;][]{rc3}, we have identified the stellar stream that we report here. The distance to NGC 3938 is not known precisely, with estimates ranging from 13.1 Mpc based on bulk flow models \citep{ohlson} to 
17.9 Mpc based on the type II-P supernova 2005ay \citep{poz}. We opt to adopt the larger distance value because it is based on a direct distance estimator and because the galaxy appears to be associated with the Ursa Major galaxy cluster, which is measured at a distance of 18.6 Mpc 
\citep{tullypierce,mcdonald}, despite its relative isolation noted previously.


This paper is structured as follows: in \S2 we briefly discuss the data from the LIGHTS survey; in \S3 we examine the stream in detail and discuss potential inferences from the stream in \S4. We address details of our bright star subtraction in Appendix \ref{ap:starsub} and present the $r-$band image in Appendix \ref{ap:rband}. Magnitudes are provided in the AB system \citep{oke1,oke2}.

\begin{figure*}
\includegraphics[width=1\textwidth]{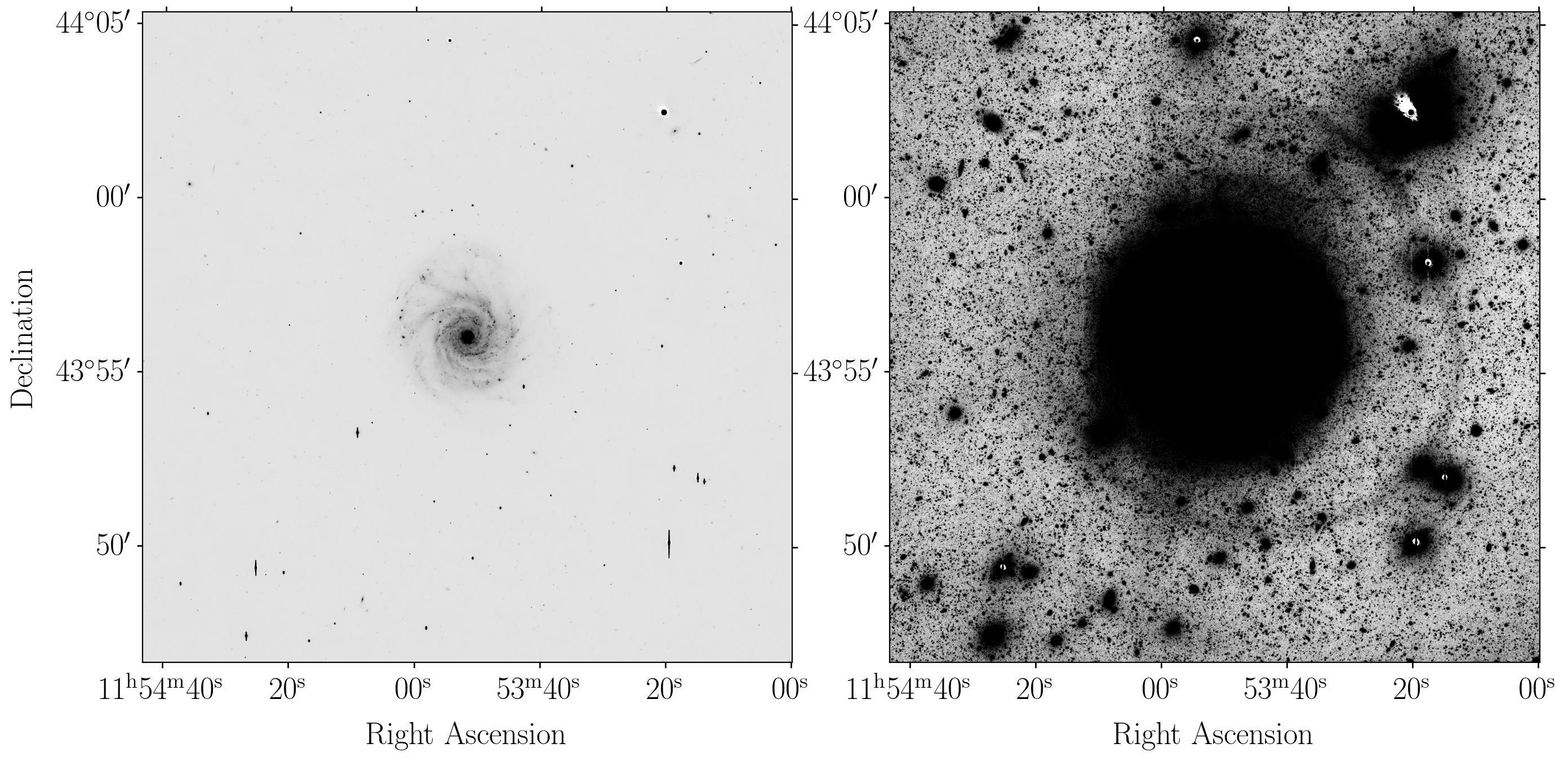}
\caption{Two views, with negative image contrast, of NGC 3938 in the $g-$band at the same angular scale but differing contrast levels. The left panel highlights the central galaxy, a classic late type disk with no clear evidence of abnormalities except for some level of asymmetry. The surface brightness apparent sensitivity in this panel is similar to that of the digitized Palomar Sky Survey image of this source. The right panel, focusing on low surface brightness features, shows the presence of a thin ($\sim 2$ kpc) tidal stream that stretches nearly 230$^\circ$ around the central galaxy. The average $g-$band surface brightness of the stellar stream is 28.7 mag arcsec$^{-2}$.}
\label{fig:n3938}
\end{figure*}

\begin{figure*}
\includegraphics[width=1\textwidth]{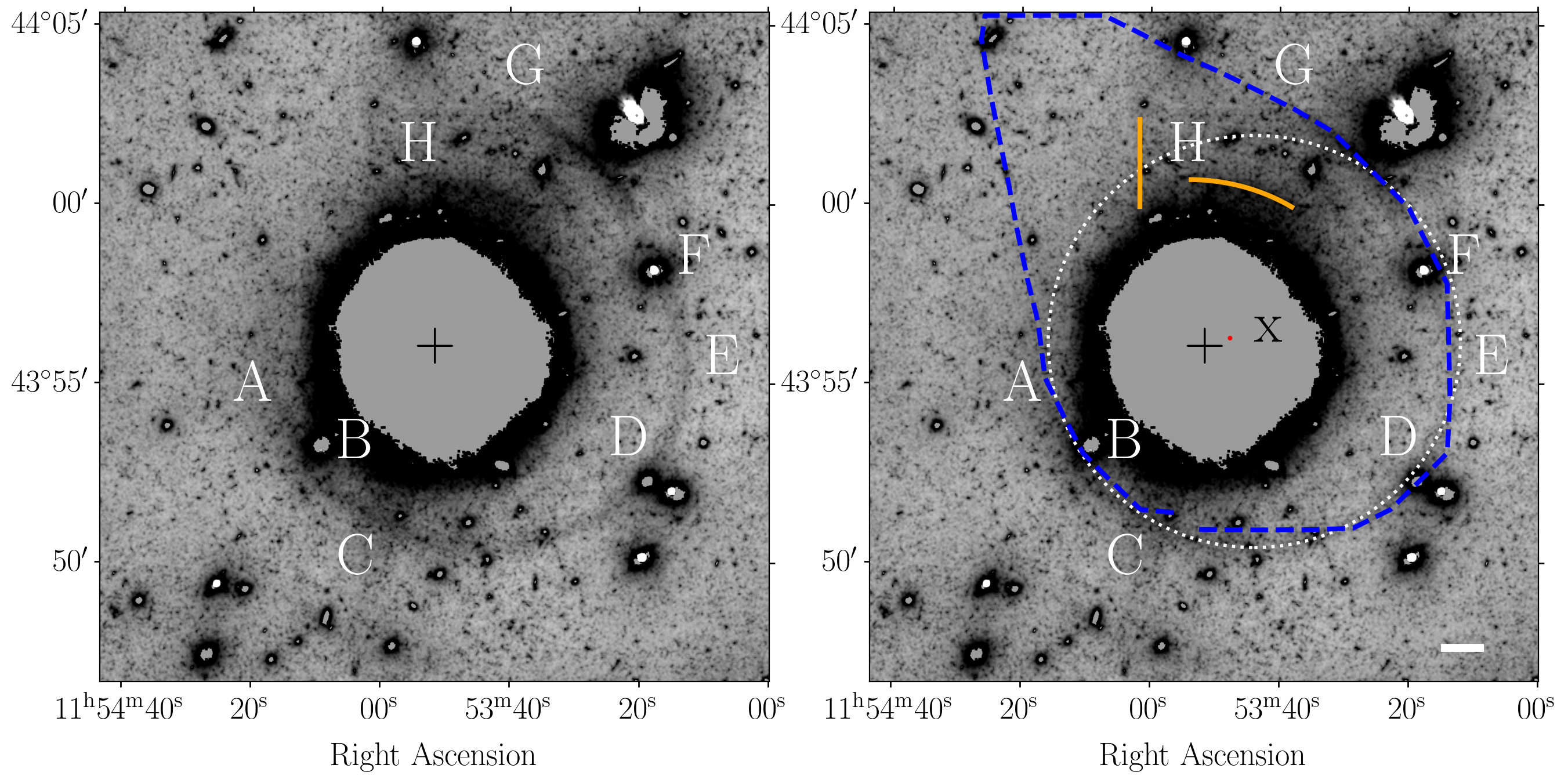}
\caption{Stellar stream with key reference points labeled.  The two panels contain the same $g$-band image, which has had some masking and smoothing applied to enhance the stream appearance. A discussion of various salient features is presented in the text. In the left panel we present the labeled locations that are referenced in the text, while in the right panel we add a reference circle that matches the majority of the stream for discussion. In orange we mark two features near location H that are discussed in the text. The location of the galaxy nucleus is marked $+$. The center of the drawn circle is marked X. Within the blue dashed line we enclose the allowed locations of the center of mass as determined from our stream curvature analysis (see \S\ref{sec:discussion}). The white scale bar at the bottom right represents 5 kpc at the distance of NGC 3938.}
\label{fig:n3938_streamer}
\end{figure*}
\section{The Data}
\label{sec:stream}

We observe LIGHTS targets with the Large Binocular Camera \citep[LBC;][]{lbc} on the LBT using the red and blue channels mounted simultaneously on each focus of the twin telescope.
Each LBC camera contains four CCDs, with a pixel scale of 0.224 $\arcsec$ pixel$^{-1}$ for a total field of view for each LBC camera of roughly 23$\arcmin$ $\times$ 25$\arcmin$. We used the g-SLOAN filter on the blue camera and the r-SLOAN filter on the red camera. 
Tailored observing and data reduction procedures are required to address different and
complex observational challenges such as scattered light, strongly saturated stars,
and ghosts. These and other issues are described in detail in \cite{lights-z}, \cite{sedighe} and \cite{golini}. NGC 3938 is one of the few additional galaxies that satisfies the selection criteria presented by \cite{lights-z} for which we have obtained data since that publication and now include in the LIGHTS sample. For comparison to the other galaxies in the sample, the $g$ and $r-$band  absolute magnitudes for NGC 3938 are $-20.4$ and $-21.0$ mag, respectively, and $E(B-V)$ is estimated to be 0.021 mag (using the \cite{sfd} maps). The data collection and reduction procedures are the same. The resulting limiting surface brightnesses are $\mu_g = 31.22$ mag arcsec$^{-2}$ and $\mu_r = 30.42$ mag arcsec$^{-2}$ (3$\sigma$ limit in an area equivalent to a 10\arsec $\times$ 10\arsec\ box), almost a magnitude fainter in $g$ than, and comparable in $r$ to, the anticipated depth of the Rubin/LSST stack.

We take an additional step not described in the existing LIGHTS survey references that involves removing as much of the extended scattered light from bright stars as possible. We use the LIGHTS LBT point spread function presented by \cite{sedighe} to subtract the scattered light of stars with $mag_{G}<13.5$ \citep[G-mag in Gaia DR3 catalog; ][]{gaia} in the individual frames prior to the sky subtraction and mosaicing process (for details see Appendix~\ref{ap:starsub}). This is part of a continuing effort to further improve the LIGHTS data processing.

We present the resulting $g-$band image of the NGC 3938 field in Figure \ref{fig:n3938} at two different intensity levels. The left panel shows the nature of central galaxy, a late-type disk galaxy seen nearly face-on with little if any signs of disturbance and no evident close companions. The right panel shows the presence of a thin tidal stream that extends more than half way around the galaxy and evidence for more broadly distributed tidal material, including various shell like structures at smaller radii than the stream to the northwest and a radial spur to the southeast. The complementary $r-$band image is presented in Appendix \ref{ap:rband}.

Given the apparent face-on orientation of both the disk and the stream, there is the possibility that the stream is an outer disk feature rather than an independent structure. Outer disk features include the strings of outer disk star forming regions \citep{thilker,hf} and the ``feathers" arising from tidal interactions noted by \cite{martinez-delgado23}. The feathers are material drawn from the disk and there is no such connection visible between the disk of NGC 3938 and the structure we are discussing. Furthermore, there is no evidence of star formation in the form of tight stellar knots along the structure. As such, we conclude that what we have found is most likely a stellar stream rather than an outer disk feature.

\section{The Stellar Stream}
\label{subsec:image processing}

At first glance, the stream appears to be quite regular, but upon further inspection, we find a number of irregularities that, in principle, offer a set of features that will constrain dynamical models. After masking bright sources, we measure a mean surface brightness for the stream of
approximately 28.7 mag arcsec$^{-2}$ and 28.1 mag arcsec$^{-2}$ in $g$ and $r$, respectively. These values are uncertain due to the treatment of masking, but differences when two of us (GG and IRC) do the masking and photometry independently are small ($<$ 0.1 mag) except along the brightest portion of the stream where values differ instead by a few tenths of magnitude.  In the $g-$band, this surface brightness is $\sim 2$ mag arcsec$^{-2}$ fainter than the mean value of streams identified using the Legacy Survey images \citep{sola2025}, but well within the anticipated reach of the Rubin/LSST stack. The surface brightness along the stream varies by as much as $\sim$ 1 mag arcsec$^{-2}$.

To gain an understanding of the stream's possible progenitor, we proceed to estimate the stellar mass of the stream. 
Within a visually-defined aperture that traces the stream, we measure a  mean $g-r$ color of 0.57,
adopt a standard \citep{roediger} mass-to-light ratio for this color for a Chabrier \citep{chabrier} initial mass function (IMF) of 1.5, and measure the stream luminosity ($2.5\times10^7$ L$_\odot$), to infer a progenitor mass of $\sim 3.7\times 10^7$ M$_\odot$.
We measure the luminosity by masking bright localized sources and replacing those pixels with the median value of the unmasked pixels after 3$\sigma$ rejection. To estimate the maximum uncertainty of this measurement, we repeat the analysis using either no masking or masking without replacing the values in those masked pixels. These provide overly conservative upper and lower values of the stellar mass that are roughly a factor of 2 higher and lower than our preferred estimate (in comparison to the $\sim$ 10\% changes seen with the moderate mask changes of the two independent efforts). 

For context, in terms of Local Group galaxies, the mass we derive lies between that of the Fornax dSph and NGC 185 \citep{mac12,kirby}.
Furthermore, our stream's stellar mass is roughly an order of magnitude lower than those of streams identified from currently-available wide-field surveys \citep[e.g., median mass $= 6.2\times 10^8$ M$_\odot$;][]{sola2025}, but similar to the extreme Giant Coma Stream \citep[M$_* = 6.8\times10^7$ M$_\odot$ and similar surface brightness;][]{roman}. Note that \cite{sola2025} adopt an approach that uses a Salpeter IMF, but this difference results in mass estimates that are only $\sim$ 0.25 dex larger than when using a Chabrier IMF \citep{bernardi}.
We have not corrected for Galactic extinction, but the extinction in this field is low, $E(B-V) = 0.021$ mag.

We begin our description of the stream near the location labeled A in Figure \ref{fig:n3938_streamer}, which marks the start of the detected stream if we proceed counterclockwise around NGC 3938. We visually trace the beginning of the stream to a position slightly to the left of the crossbar in the letter A. Relative to the nucleus of NGC 3938, which we label in Figure \ref{fig:n3938_streamer} with the plus sign, the start of the stream is at an angle of 103$^\circ$ (measured from north to east) and at a projected distance of 25.1 kpc. 

We trace the stream from A toward B, although the bright star near location B obscures the exact track.  There is an inflection in the stream as it continues beyond B. 

There are two features that we highlight associated with location C. First, there is a radial, broad spur of low surface brightness light crossing the stream nearly perpendicularly. Coincidentally with the crossing of the spur and the stream, the stream appears to bend slightly inward (toward NGC 3938). We speculate that the morphology of the stream at this location is related to the influence of the progenitor of the spur, and we will discuss this further when discussing location H. The stream then proceeds smoothly to location D.

We highlight location D because it lies roughly halfway around the stream and there is a bright source that coincides with the stream. It might be tempting to hypothesize that this source is the remaining core of the disrupted progenitor, but this source is like a star. The object is both slightly offset laterally from the stream and shows a bleed trail, both of which support this interpretation. Unfortunately, this star may obscure what might remain of the actual progenitor core.

The stream kinks slightly at location E. Bends in the stream are valuable constraints on the shape of the underlying gravitational potential \citep{nibauer}. In this case, there is a possibility that the visual impression of a kink is influenced by an unfortunate superposition of faint foreground stars, but the change in curvature appears to be consistent from locations E to F. Another way to see the change in curvature is to compare to a circle drawn to coincide with the stream so far (right panel, Figure \ref{fig:n3938_streamer}). After location E, the stream is clearly interior to the circle. We conclude that the stream changes curvature near location E.

Unfortunately, there is another bright star superposed on the stream at location F. It lies at a position along the stream that appears to be less clearly defined than other portions of the stream. Density fluctuations along the stream are potentially valuable diagnostics that can highlight recent interactions with dark matter substructure, although in certain cases they can be the result of the epicyclic motions of the escaping stars from the progenitor causing imperfect phase mixing \citep{kupper}. Unfortunately,   in this case, at this location, the stream density is obscured by a bright Galactic star.

Between locations F and G we find the highest surface brightness section of the stream, with mean surface brightness within this feature of $\sim$ 27.6 and 27.1 mag arcsec$^{-2}$ in $g$ and $r$, respectively. This would naturally be associated with the disrupted body of the progenitor but is surprisingly located near one end of the stream. The stream in this section shows a distinct tilt, which may reflect the split between leading and trailing tails, assuming that this high surface brightness segment of the stream indicates the location of the last remnant of the progenitor. There is no evidence of a central, surviving core. Alternatively, this feature is unassociated with the stream and is instead a shell arising from the disruption of a different satellite, perhaps that associated with the broad radial feature identified near location C. 

To further examine this last possibility, we measure the color of the stream, masking stars and correcting for Galactic extinction. Two of us (GG and IRC) performed the masking and photometry independently. The resulting surface brightness and color measurements are consistent between the two efforts, except for in the stream region between F and G where the surface brightnesses differ by a few tenths of a magnitude. The color of the feature between F and G, $\sim 0.6$ mag, is consistent between the two efforts and with the colors measured at various locations along the stream, although there are variations along the stream that are  $\gtrsim 0.5$ mag. The largest of these variations are presumably due to poor masking, but these variations make it difficult to reach a conclusion regarding the origin of the feature between F and G on the basis of its measured color. There is no evidence based on its color that it might belong to a different progenitor.

We cannot trace the stream much beyond the location G. We place the end, below and just to the left of the letter G, at an angle of 337$^\circ$ and a distance of 36.3 kpc. Therefore, the stream traces a curve that extends $\sim$ 234$^\circ$ around NGC 3938 and is not centered on the galaxy. Its mean distance from the host ($\sim 32.8$ kpc) is comparable to that of some of the streams identified by the STRRING survey \citep{sola2025}.

The final location we label is H. There are two features to discuss near this location. First, there is a vertical low surface brightness feature to the left of the letter H. Because it is exactly vertical, we attribute this to a slight background matching error when mosaicing the various images used to create this image. Second, there is a surface brightness truncation in the body of NGC 3938 just slightly below the letter H, extending clockwise for about 30$^\circ$. This shell is nearly antipodal with the radial low surface brightness spur noted at location C. The progenitor of these two structures appears to have been on a nearly perfectly radial orbit and seems to have suffered instantaneous disruption because only one shell is visible, unless the high surface brightness feature at location G is an associated shell. There is also the possibility that additional shells are obscured by the galaxy disk outskirts. We suggest that, prior to its disruption, this satellite did influence the morphology of the stream near position C. If so, this connection may provide geometric and timing constraints on any full model of the NGC 3938 system.

\section{Discussion}
\label{sec:discussion}

The stream from location B to E is nearly circular, with an inferred center that is visually offset from the nucleus of the galaxy (right panel, Figure \ref{fig:n3938_streamer}). If one could confirm that this offset, which corresponds to $\sim$7 kpc, reflects an offset of the gravitational potential from the center of the galaxy, then there are several key implications. First, this system would be an example of the sloshing of a galaxy within the potential that has been hypothesized to occur due to interactions and would lead to a lopsided appearance of the disk \citep{jog,kornreich}.
Many galaxies have stellar disks that are lopsided \citep{rz} and NGC 3938, at its outer isophotes (Figure \ref{fig:n3938_streamer}, right panel) is lopsided. 
Second, the quantitative degree of this sloshing could be used to constrain the steepness (or cuspyness) of the dark matter density profile, which in turn can constrain the nature of the dark matter particle. Standard models suggest that offsets should be common but an order of magnitude smaller than inferred here \citep{kuhlen}. Finally, the offset itself would be direct evidence for the physical reality of dark matter, using much the same argument as presented for the Bullet Cluster \citep{clowe}, where there is a manifestation of a separation between the centroid of the observed baryonic matter and the inferred dark matter.

Unfortunately, in reality, the stream is not quite circular. More complex orbits, such as rosette orbits, can appear to be miscentered if only a portion of the orbit is visible \citep[for examples see Figure 4 of][]{buist15}. Finally, the stream does not represent the orbit of any individual object \citep{sanders}, but rather snapshots of orbits of objects with a range of different energies, so care is warranted, and we cannot simply conclude that the galaxy and gravitational potential are offset. 

\begin{figure*}
\begin{center}
\includegraphics[width=0.9\textwidth]{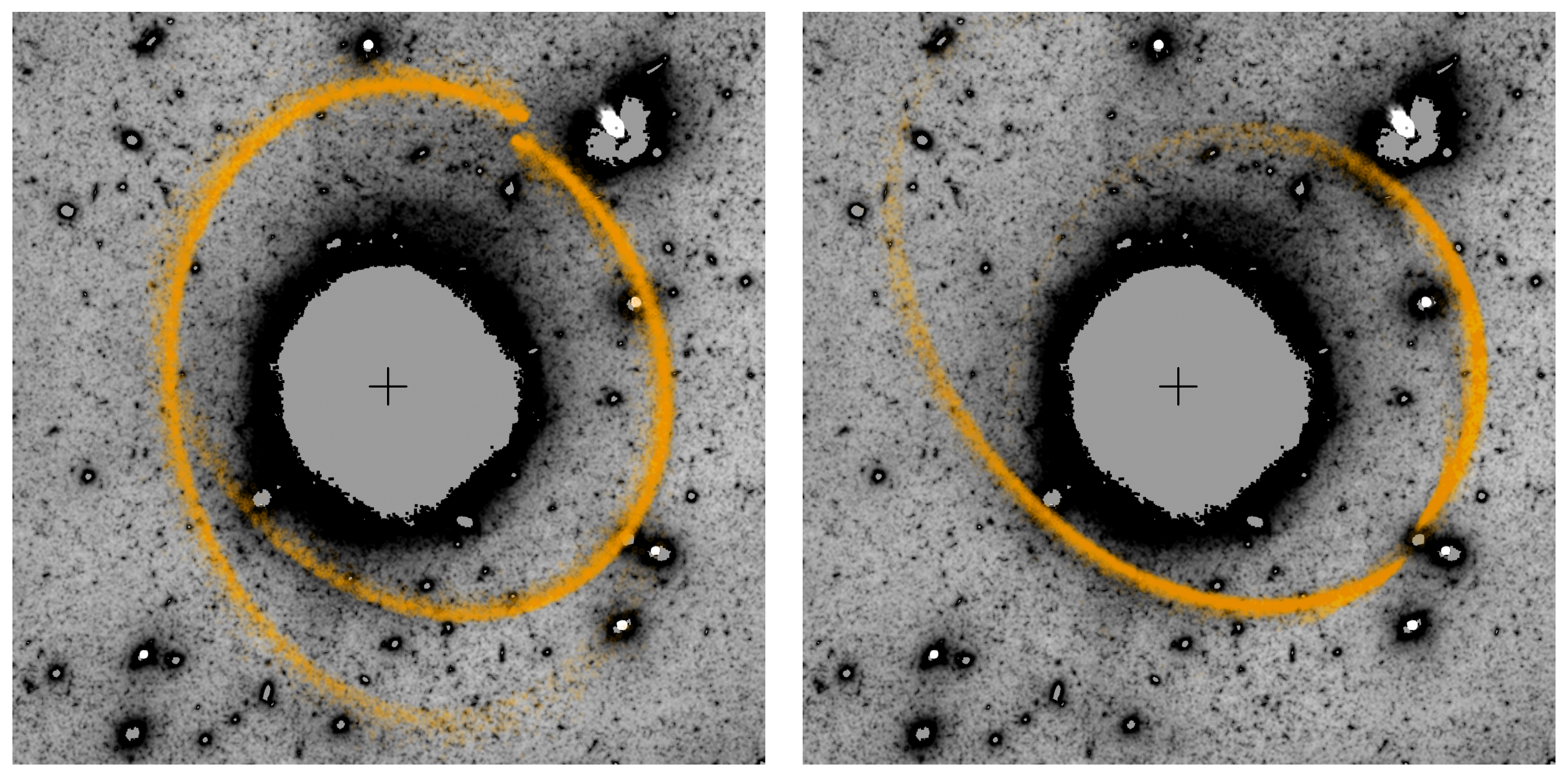}
\end{center}
\caption{Example particle spray stream orbits in orange superposed on the LIGHTS image. In the left panel we present one possible configuration where the progenitor is located at the top of the observed stream, near the highest surface density region (between locations G and F). In the right panel we present a second possible configuration where the progenitor is located near the midpoint of the observed stream (location D).}
\label{fig:simulations}
\end{figure*}

We present a preliminary analysis using the stream curvature modeling technique described in detail by \cite{nibauer}. In summary, by analyzing the curvature of the stream’s track, it is possible to constrain the shape of the underlying gravitational potential without making assumptions about its normalization. As shown in \cite{nibauer}, streams that closely follow great circles about their host are not informative in determining the halo shape. The curvature method can also be used to constrain the center of mass of the host galaxy. In this work, we focus on the constraints on the center of mass by first annotating the stream track with a hand-drawn spline and then differentiating it twice to obtain the curvature.

The result we draw from an application of that analysis is that the center of the potential cannot be constrained to lie off the galaxy center. In other words, we find no evidence that the center of light differs from the center of mass, despite the off-center appearance of the stream.
The center-of-mass locations allowed are shown in Figure \ref{fig:n3938_streamer}. Although streams resembling great circles are not found to be constraining for the flattening of the halo, constraints on the halo mass and radial profile can be derived from generative modeling \citep{nibpear2025}. We defer a full parameter exploration of this stream with generative models to future work.

We now use a particle spray approach \citep{fardal} to test whether simple generative models can reproduce the morphology of the observed tidal stream. We perform simulations using the \texttt{streamsculptor} python package \citep{Nibauer25}.
Using models of different progenitors in a spherical NFW halo \citep{nfw} with a Miyamoto-Nagai disk \citep{miyamoto}, and placing the progenitor either at the most overdense region of the stream (between locations F and G in Figure \ref{fig:n3938_streamer}) or near the midpoint of the stream (location D), we easily find models that reproduce the general stream morphology, although each has shortcomings (see below). The halo in these models has a scale mass of $\sim 10^{12}$ M$_\odot$ and a scale radius of $\sim$ 20 kpc, while the disk has a mass of $5 \times 10^{10}$ M$_\odot$, a scale length of 3 kpc, and a scale height of 0.2 kpc. These values are not derived from a full parameter study nor is the matching between theory and observation optimized in a quantitative manner. Our aim is simply to demonstrate that streams that are similar in appearance to that observed can be generated without resorting to unusual circumstances such as an offset between the galaxy and the global potential. 
The simulations are typically insensitive to the parameters of the disk, within plausible values.

In Figure \ref{fig:simulations} we show the particle distributions in two of our particle spray simulations. To generate these examples, we consider the two sky positions mentioned previously for the progenitor ($M_{\rm prog} =5\times10^7~M_\odot$) and launch orbits over a grid of velocities and directions to visually compare the generated streams to the observations. Because the internal structure of the satellite progenitor is not modeled, the small-scale density structure of the simulated stream is not expected to be accurate. However, the tidal debris track is expected to match more detailed $N-$body modeling (e.g., \citealt{Pearson}). 
In a model where the progenitor remnant is placed at the densest part of the stream, the model matches the observed stream but also suggests that we should find a leading stream (or trailing stream, depending on the unknown sense of the orbital motion) that we do not find. Asymmetry in the leading vs. trailing stream prominence can arise due to details of the mass loss that are more complex than in the simple picture of tidal mass loss \citep{choi,varisco} and in exotic dark-matter models \citep[e.g.,][]{kesden}, although in neither case is there a demonstration of an asymmetry as large as would be needed here.

In a model where the progenitor is centered near the midpoint of the stream, the match is better in the sense that it does not predict a stream where we do not find a stream, although we are left to explain the higher stellar density at the northwestern tip of the stream. Perhaps, as noted previously, this feature is unassociated with this stream. We conclude that it is possible to match the general morphology of the observed stream with a standard galaxy and dark matter halo model that are co-centered, although details of the density structure, such as surface brightness variations and apparent kinks, remain unresolved. No higher degree of complexity is required to match the bulk properties of the stream, including the apparent miscentering on the galaxy,  although these models do not preclude such an offset.

To establish an offset, it may be necessary to model jointly several aspects of the system, including the lopsided disk, the disrupted system that produces the spur and possible shells, and the stream that we have identified. It is intriguing that the lopsidedness extends in the direction of the apparent center of the stream orbit. We have confirmed from {\it Spitzer} 3.6$\mu$m imaging \citep{s4g} that the lopsided feature is a smooth distribution of stars rather than an outer disk spiral arm.
It may be that only with a model of the entire system will there be enough information to conclusively determine if there is an offset of the global potential center from that of the galaxy. Nevertheless, this discussion shows how even the modeling of a relatively simple system in appearance can quickly demand a highly complex model.

One final aspect to discuss is the possible inclination of the stream to the line of sight. If the stream is roughly circular in cross-section, inclining it would result in surface brightness enhancements that vary azimuthally. In principle, this effect might provide an opportunity to constrain the inclination, but in practice surface variations arising from other factors, some discussed previously, complicate the situation.
\section{Summary}
\label{sec:summary}

It is anticipated \citep{martin} that most (80\%) images of galaxies that reach a surface brightness limit of $\sim$30.5 mag arcsec$^{-2}$ will show signs of previous interactions. Indeed, we show here one example of such an image from the LIGHTS program that contains an intriguing stellar stream.
The stream is thin ($\sim 2$ kpc), appears to lie nearly on the plane of the sky, and wraps more than half way around a host galaxy that is otherwise apparently isolated. We estimate that the stellar mass of the progenitor was $\sim  3.7\times 10^7$ M$_\odot$. 

With preliminary models, we find that the bulk properties of the stream, including its apparent miscentering on NGC 3938, can be reproduced with standard assumptions regarding the host galaxy. It remains to be seen what constraints can be derived from a fully self-consistent model of the interaction, including a possible interaction with a second, radially-orbiting, disrupted satellite. Overall, this system is likely to be one of the simplest to model and, as such, a good test of what information can be robustly recovered from the 2-D projection of galactic tidal debris. The requirements for a complete model quickly escalate to a high degree of complexity. Nevertheless, the expected bounty of stellar streams from upcoming surveys should provide numerous comparable examples to the stream presented here and justifies further development of the analysis methodology. 

\section*{Acknowledgments}

The LBTO is an international collaboration among institutions in the United States, Germany, and Italy. Observations have benefited from the use of ALTA Center (alta.arcetri.inaf.it) forecasts performed with the Astro-Meso-Nh model. Initialization data of the ALTA automatic forecast system come from the General Circulation Model (HRES) of the European Centre for Medium Range Weather Forecasts.

We thank the anonymous referee for a thorough reading and helpful comments that improved the presentation.
DZ gratefully acknowledges support from the NSF (AST-2510821) and NASA (22-ADAP-0011).
IT acknowledges support from the State Research Agency (AEI-MCINN) of the Spanish Ministry of Science and Innovation under the grant PID2022-140869NB-I00 and IAC project P/302302, financed by the Ministry of Science and Innovation, through the State Budget and by the Canary Islands Department of Economy, Knowledge, and Employment, through the Regional Budget of the Autonomous Community. We also acknowledge support from the European Union through the following grants: "UNDARK" and "Excellence in Galaxies - Twinning the IAC" of the EU Horizon Europe Widening Actions programs (project numbers 101159929 and 101158446). Funding for this work/research was provided by the European Union (MSCA EDUCADO, GA 101119830). GG acknowledges support from the PID2022-140869NB-I00 grant from the Spanish Ministry of Science and Innovation.
M.N.Le is funded by the European Union (MSCA EDUCADO, GA 101119830). RIS acknowledges the funding by Governments of Spain and Arag\'on through FITE and Science Ministry (PGC2018-097585-B-C21, PID2021-124918NA-C43). 
MM acknowledges support from grant RYC2022-036949-I financed by the MICIU/AEI/10.13039/501100011033 and by ESF+, grant CNS2024-154592 (CUTIE) financed by 
MICIU/AEI/10.13039/501100011033 and program Unidad de Excelencia Mar\'{i}a de Maeztu CEX2020-001058-M.
JR acknowledges financial support from the Spanish Ministry of Science
and Innovation through the project PID2022-138896NB-C55
IRC acknowledges support from grant PID2022-140869NB-I00 from the Spanish Ministry of Science and Innovation.
GM acknowledges support from the UK STFC under grant ST/X000982/1.
This work was supported by a research grant (VIL53081) from VILLUM FONDEN and by the European Union (ERC, BeyondSTREAMS, 101115754) grant. 
SR acknowledges support from the GEELSBE2 project with reference PID2023-150393NB-I00 funded by MCIU/AEI/10.13039/501100011033 and the FSE+, and also the Consolidación Investigadora IGADLE project with reference CNS2024-154572. SR also acknowledges support from the project PID2020-116188GA-I00, funded by MICIU/AEI/10.13039/501100011033. SR gratefully acknowledges the financial support of the Department of Education, Junta de Castilla y León, and FEDER Funds (Reference: CLU-2023-1-05). JR acknowledges financial support from Plan Propio de Investigación 2025 submodalidad 2.3 of the University of Cordoba. IRC and SGA acknowledge support from grant PID2022-140869NB-I00 from the Spanish Ministry of Science and Innovation. Views and opinions expressed are however those of the author(s) only and do not necessarily reflect those of the European Union, the European Research Council, or any other funding agency. Neither the European Union nor the granting authority can be held responsible for them.

\software{
Astropy              \citep{astropy1, astropy2},
Gnuastro             \citep{gnuastro,gnuastro2},
Matplotlib           \citep{matplotlib},
NumPy                \citep{numpy},
pandas               \citep{pandas},
SciPy                \citep{scipy1, scipy2},
streamsculptor       \citep{Nibauer25}
}

\bibliography{references.bib}{}
\bibliographystyle{aasjournal}
\appendix
\section{Star subtraction on individual frames}\label{ap:starsub}

As introduced in \S2, we improve on our previous LIGHTS image processing by subtracting the  scattered light from bright stars in the individual frames prior to sky subtraction and frame stacking. In this section, we provide more details of this process and our reasoning.

\begin{figure*}
\centering
\includegraphics[width=0.7\textwidth]{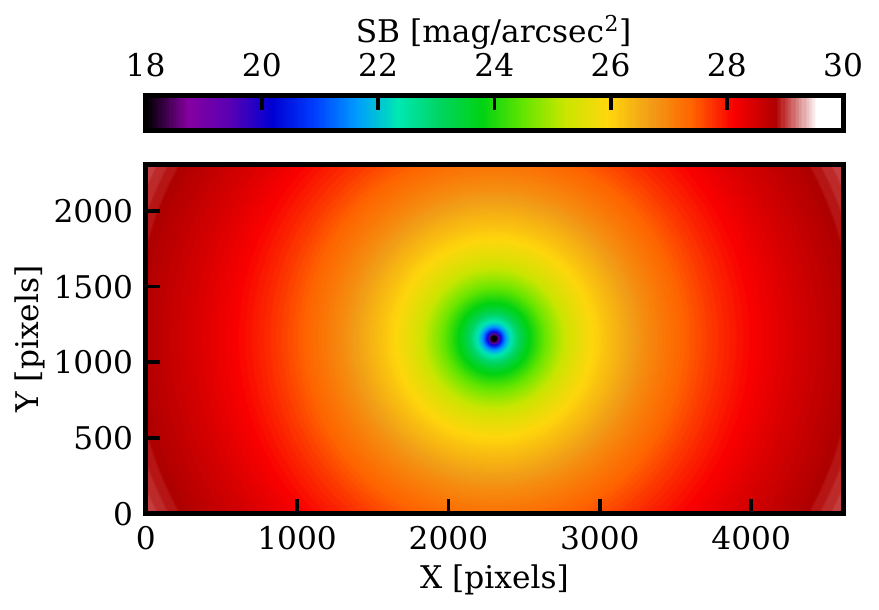}
\caption{Magnitude 8.7 star simulated with the \textit{g}-band PSF from \citet{sedighe} in a single LBT CCD. This simulation corresponds to the bright star HD 103082 in the field of NGC 3938, which has a Gaia G-mag of 8.73 mag.}
\label{fig:psf_ccd}
\end{figure*}

Previous work focusing on low surface brightness (LSB) features \citep[e.g., ][]{tf} emphasizes that a key step in a LSB-friendly reduction is the measurement and subtraction of the background sky. In LIGHTS  \citep{lights-z}, we have applied a two step process that includes 
measuring and subtracting a constant value from the images after defining a proper mask. With the mask, we attempt to cover all sources of light that are not from a uniform background, including diffuse light from galaxies and Galactic Cirrus, and the scattered light of bright stars. However, we now find that in fields like that of NGC 3938, a star can be too bright to be fully covered by the mask because its scattered light covers the full CCD, or even extends to other CCDs. In such cases, we do not measure a proper value for background subtraction (see Fig.~\ref{fig:psf_ccd}) with our original approach.

To overcome this problem, we apply a novel algorithm to subtract the scattered light of bright stars from the individual frames {\sl prior} to the background subtraction. This algorithm consists of three major steps that are done iteratively on a star-by-star basis, beginning with the brightest star and concluding once we have accounted for all stars in the field of view, and 0.2$^\circ$ outside the field, that have Gaia G-band magnitudes $<$ 13.5 mag. We apply this procedure iteratively because the light from the brighter stars can affect the fitting for the fainter ones.

\begin{enumerate}
    \item First, we identify those images where the scattered light of the star in question is significant. In this context, we define an image as a single exposure that includes all four CCDs. To assess, we calculate the radius at which the LIGHTS PSF \citep{sedighe}, scaled to the magnitude of the specific star being considered, has a surface brightness of 26.5 mag arcsec$^{-2}$. This value  is $\sim$ 50 times fainter than the typical sky measured in our images. If a circle of that radius overlaps any part of the image then we proceed to correct the image for the scattered light from that star.
    \item Second, we construct a model of the scattered light. After selecting the affected images, we measure the star's radial intensity profile. Each image consists of four CCDs and we measure the star's profile from the CCD image that has the greatest scattered light within the circle mentioned previously. We then normalize the LIGHTS PSF to this measured profile. Because this process is done prior  to background subtraction, our normalization needs to account for the background. To allow for this, we adopt $I_{star}=\alpha I_{PSF}+SKY$, where $I$ is the intenstity as a function of radius and $SKY$ is a constant, and fit using  using \texttt{Scipy}'s  \texttt{curve$\_$fit} where the data lie between the saturation level ($I_{star}\sim5\times10^{3}$ ADU) and the background level ($\sim1.1$ times the background measured on the frame prior to the star subtraction). We present an example of the results in Figure ~\ref{fig:star_prof}.

    \begin{figure*}
\begin{center}\includegraphics[width=0.6\textwidth]{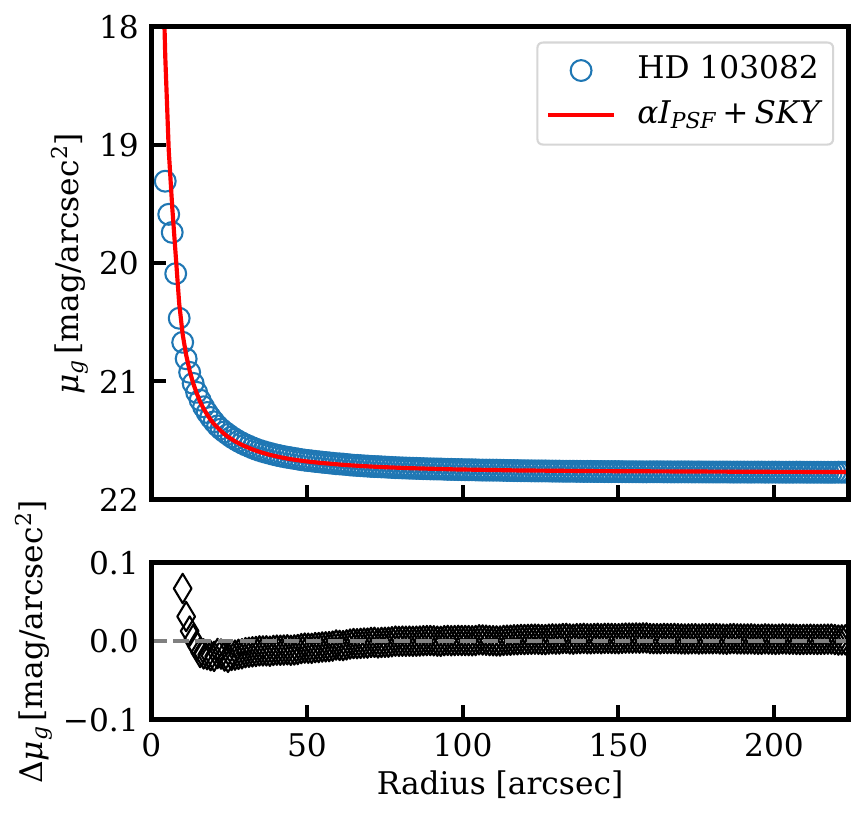}
\end{center}
    \caption{Profile of the star HD 103082 prior to the background subtraction in an individual frame, compared with the fitted values for the PSF (upper panel). The values fitted here are $\alpha=8601.75$ and $SKY=3238$ ADU. The differences between model and observations are shown in the lower panel.}
    \label{fig:star_prof}
    \end{figure*}

    \item Third, we use the $\alpha$ computed in the previous step to normalize the PSF and subtract it 
    using using \texttt{Gnuastro}'s \texttt{astscript-psf-subtract} \citep{gnuastro-psf} across all four CCDs.
 Once the scattered light of all the contaminating stars is subtracted, we compute and subtract a constant background and continue with the data reduction steps explained in \citet{lights-z}.
    We have assumed that the PSF normalization factor, $\alpha$, is the same across the four CCDs in a single frame. We argue that this is appropriate because we have previously corrected for small gain differences across the CCDs.
    
\end{enumerate}

\begin{figure*}
\includegraphics[width=1\textwidth]{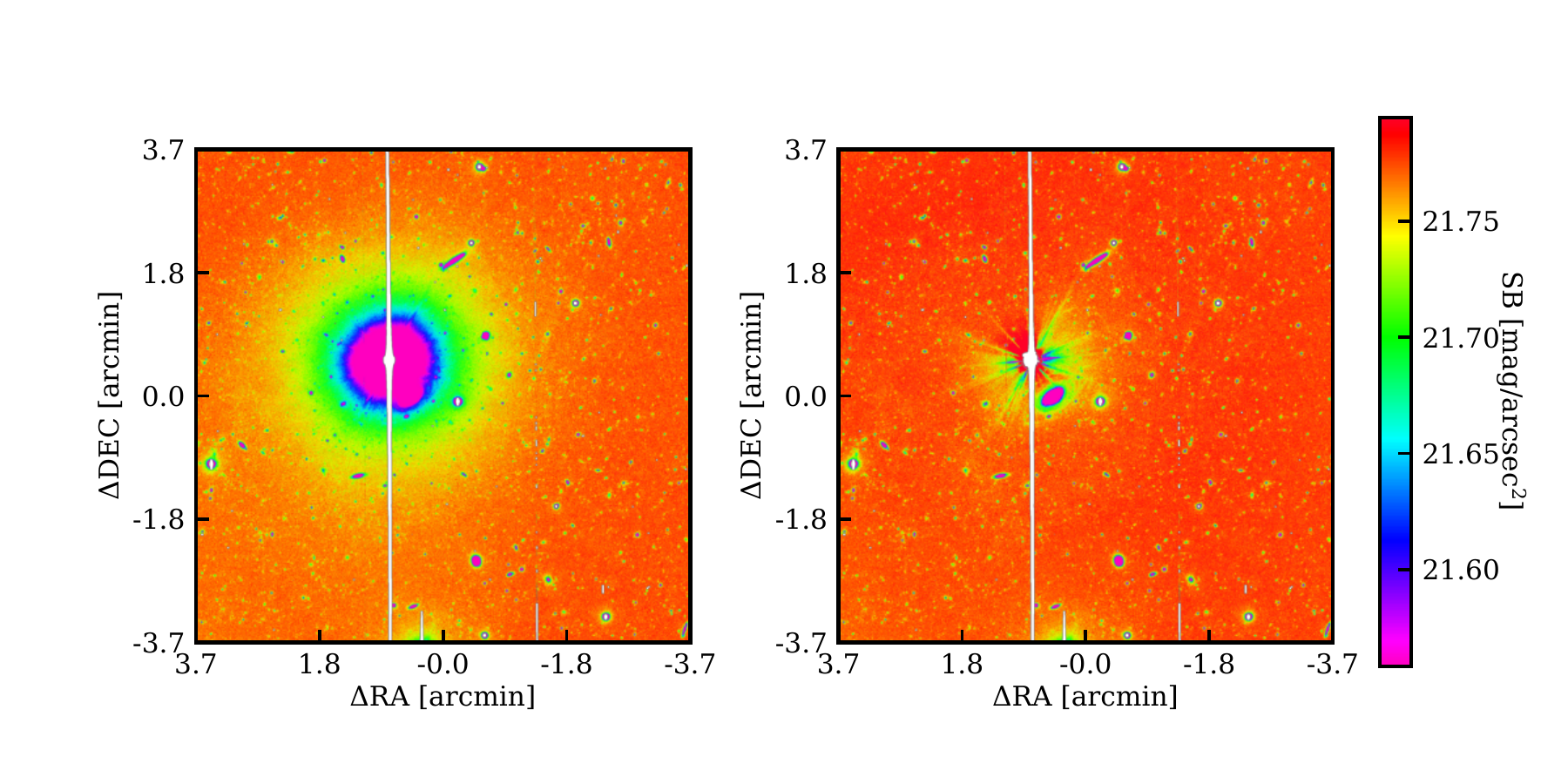}
\caption{A close-up of one individual $g$-band frame before (left) and after (right) our subtraction of the scattered light from the star HD 103082. In neither case is the mean background subtracted.}
\label{fig:example_starsub}
\end{figure*}

 In Figure ~\ref{fig:example_starsub} we present the subtraction of the star HD 103082 in the $g$-band in one of our individual frames. After removing the star, the scattered light is mitigated, the  background is more clearly visible and flatter, and new sources, such as the galaxy just below the stars, become visible. The subtraction is clearly imperfect, partly because of saturation in the very center of the stellar image and partly due to asymmetric features that are not captured in the circularized model PSF. 

\begin{figure*}
\includegraphics[width=1\textwidth]{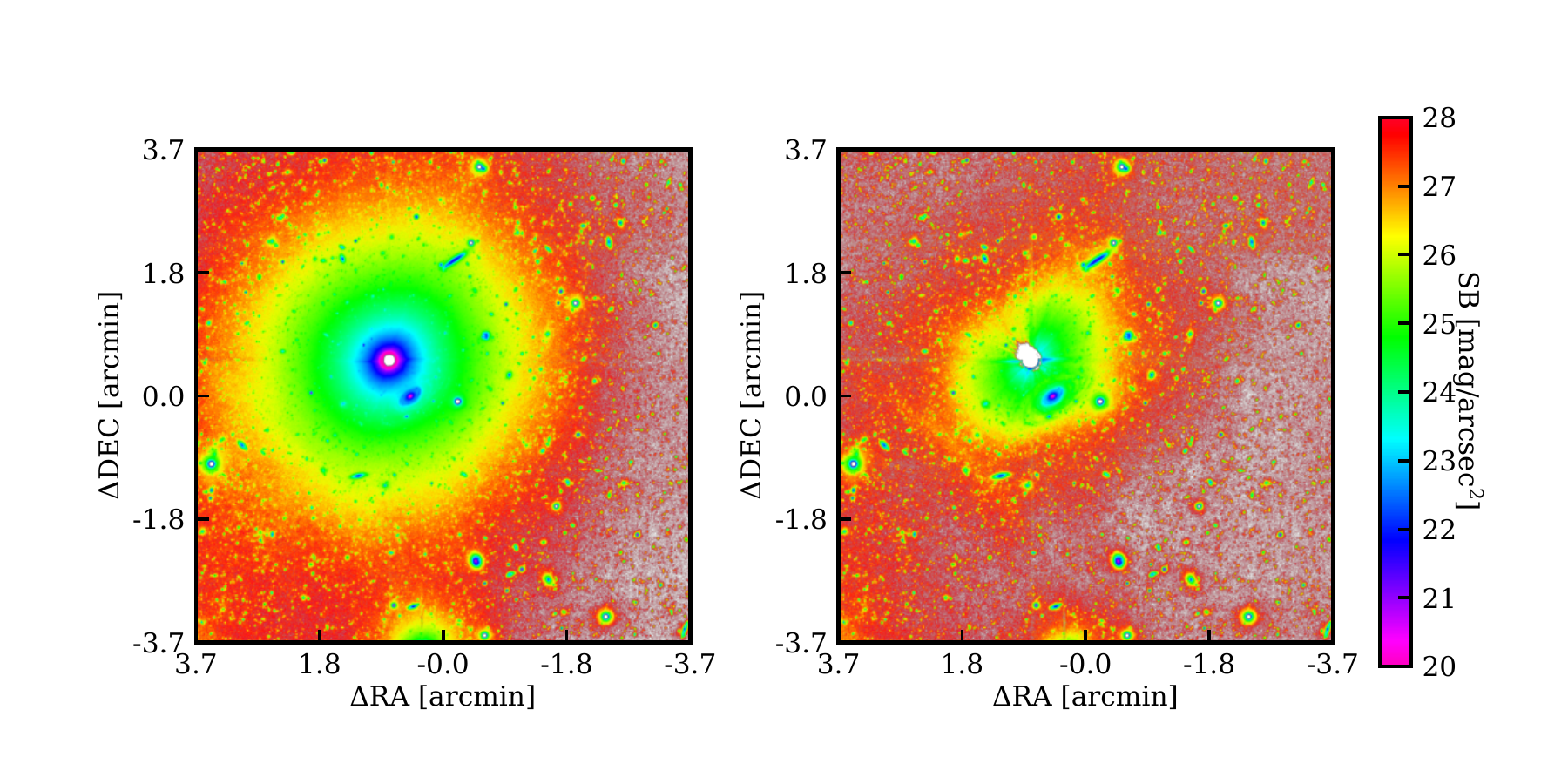}
\caption{A close-up of the final, $g$-band coadded image before (left) and after (right) subtracting the scattered light of the stars brighter than 13.5 mag (Gaia G band) in the individual frames. The nonaxisymmetric, time variable component of the PSF remains as a residual. A segment of the stellar stream becomes visible to the southeast of the bright star.}
\label{fig:coadd_starsub}
\end{figure*}

In the context of this work, the scattered light subtraction of the bright stars, especially the brightest one (HD 103082), is crucial for the stellar stream analysis. To illustrate this, we compare in Figure 
~\ref{fig:coadd_starsub} a section (same as shown in Fig~\ref{fig:example_starsub}) of the final coadded image from our original processing with the one from our improved processing. 
The improvement in the clarity of the stream between positions F and G (as labeled in Figure~\ref{fig:n3938_streamer}) is evident. 

\section{The r-band image}\label{ap:rband}
\begin{figure*}
\center{
\includegraphics[width=0.5\textwidth]{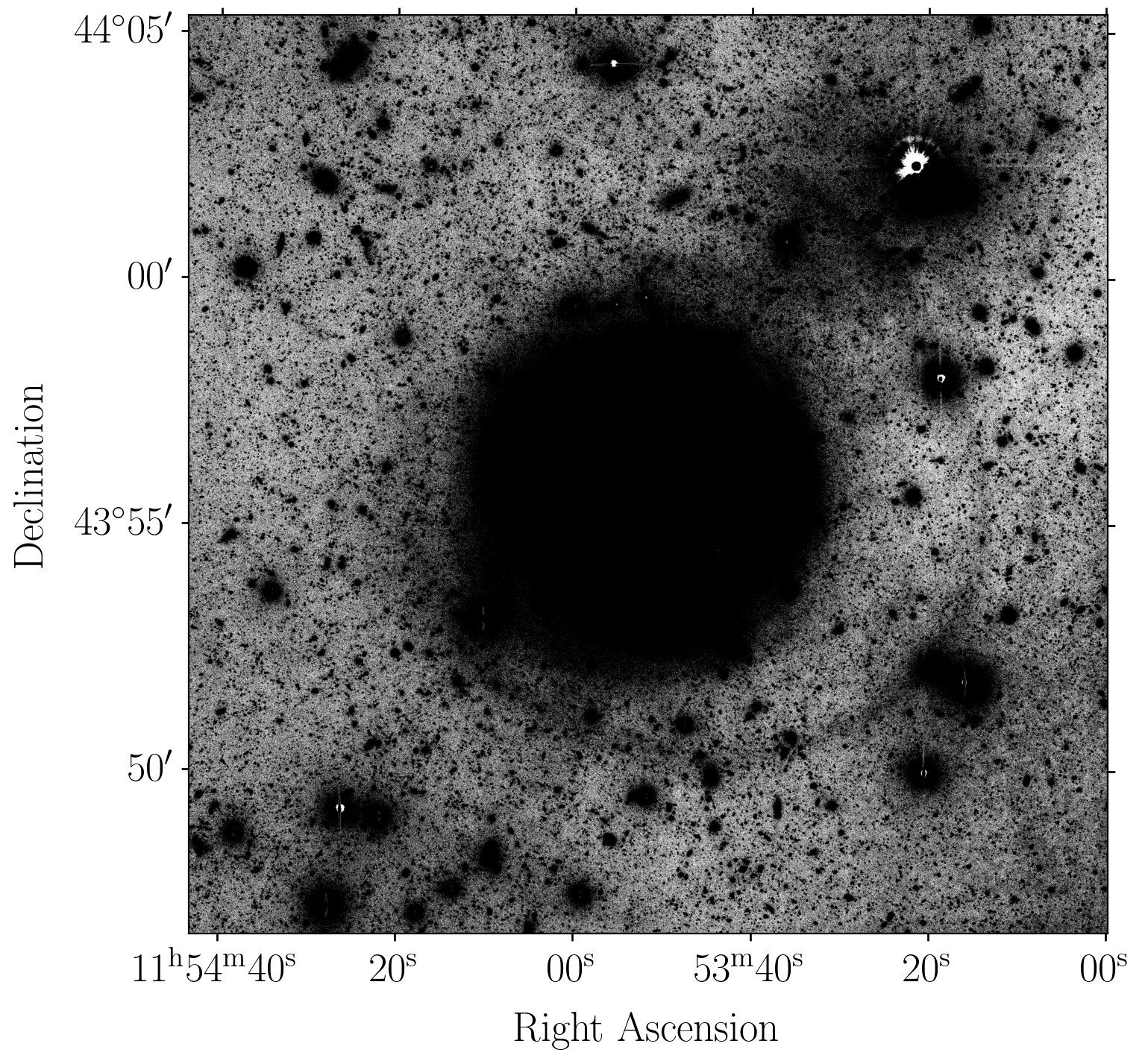}}
\caption{The $r-$band image of the NGC 3938 field and the stellar stream. }
\label{fig:rband}
\end{figure*}

We include for completeness our $r-$band image of the field in Figure \ref{fig:rband}. The stream is clearly visible in this frame as well as the various physical features we discuss in the body of the paper.

\end{document}